\pdfoutput=1
\documentclass[fleqn,usenatbib,usedcolumn]{mnras}
\usepackage[british]{babel}             
\let\la=\lesssim 
\usepackage[T1]{fontenc}                
\usepackage{graphicx}                   
\usepackage[usenames,dvipsnames]{color}

\hypersetup{pdfauthor={I. Heywood},
               pdftitle={ATCA detections of massive molecular gas reservoirs in dusty, high-z radio galaxies},
               pdfkeywords={galaxies: high-redshift -- radio lines: galaxies -- radio continuum: galaxies},
               bookmarksnumbered=true}

\volume{{\rm in press}}   

\usepackage{graphicx}	
\usepackage{amsmath}	
\usepackage{amssymb}	
\usepackage{multicol}        
\usepackage{bm}		
\usepackage{pdflscape}	
\usepackage{color}

\usepackage[T1]{fontenc}
\usepackage{ae,aecompl}

\usepackage{txfonts}

\title[Molecular gas in dusty radio galaxies at high-z]{ATCA detections of massive molecular gas reservoirs in dusty, high-z radio galaxies}

\author[Heywood et al.]
{\parbox{\textwidth}{
\begin{flushleft}
I.~Heywood$^{1,2}$\thanks{E-mail: \href{mailto:ian.heywood@csiro.au}{ian.heywood@csiro.au}}, 
Y.~Contreras$^{3}$, 
D.~J.~B.~Smith$^{4}$,
A.~Cooray$^{5}$,
L.~Dunne$^{6,7}$,
L.~G\'{o}mez$^{1,8,9}$,
E.~Ibar$^{10}$,
R.~J.~Ivison$^{11, 6}$,
M.~J.~Jarvis$^{12,13}$,
M.~J.~Micha{\l}owski$^{6,14}$,
D.~A.~Riechers$^{15}$
and P.~van der Werf$^{3}$
\end{flushleft}}
\\
$^{1}$CSIRO Astronomy and Space Science, P.O. Box 76, Epping, NSW 1710, Australia\\
$^{2}$Department of Physics and Electronics, Rhodes University, PO Box 94, Grahamstown, 6140, South Africa\\
$^{3}$Leiden Observatory, Leiden University, PO Box 9513, NL-2300 RA Leiden, the Netherlands\\
$^{4}$Centre for Astrophysics, Science and Technology Research Institute, University of Hertfordshire, Hatfield, Herts AL10 9AB, UK\\
$^{5}$Department of Physics \& Astronomy, University of California, Irvine, CA 92697, USA\\
$^{6}$Institute for Astronomy, University of Edinburgh, Royal Observatory, Blackford Hill, Edinburgh, EH9 3HJ, UK\\
$^{7}$School of Physics and Astronomy, Cardiff University, Queens Buildings, The Parade, Cardiff, CF24 3AA, UK\\
$^{8}$Departamento de Astronom\'{i}a, Universidad de Chile, Camino El Observatorio 1515, Las Condes, Santiago, Chile\\
$^{9}$Joint ALMA Observatory, Alonso de C\'{o}rdova 3107, Vitacura, Santiago, Chile\\
$^{10}$Instituto de F\'{i}sica y Astronom\'{i}a, Universidad de Valpara\'{i}so, Avda. Gran Breta\~na 1111, Valpara\'{i}so, Chile\\
$^{11}$European Southern Observatory, Karl-Schwarzschild-Strasse 2, 85748 Garching bei M\"{u}nchen, Germany\\
$^{12}$Astrophysics, Department of Physics, Keble Road, Oxford, OX1 3RH, UK\\ 
$^{13}$Physics Department, University of the Western Cape, Private Bag X17, Bellville 7535, South Africa\\
$^{14}$Scottish Universities Physics Alliance\\
$^{15}$Department of Astronomy, Cornell University, 220 Space Sciences Building, Ithaca, NY 14853, USA\\
}
\date{Last updated 20XX Jan 01; in original form 20XX Jan 01}

\pubyear{2016}

\begin{document}
\label{firstpage}
\pagerange{\pageref{firstpage}--\pageref{lastpage}}
\maketitle

\begin{abstract}

Observations using the 7~mm receiver system on the Australia Telescope Compact Array have revealed large reservoirs of molecular gas in two high-redshift radio galaxies: HATLAS\,J090426.9+015448 ($z$~=~2.37) and HATLAS\,J140930.4+003803 ($z$~=~2.04). Optically the targets are very faint, and spectroscopy classifies them as narrow-line radio galaxies. In addition to harbouring an active galactic nucleus the targets share many characteristics of sub-mm galaxies. Far-infrared data from \emph{Herschel}-ATLAS suggest high levels of dust ($>$10$^{9}$ M$_{\odot}$) and a correspondingly large amount of obscured star formation ($\sim$1000~M$_{\odot}$ / yr). The molecular gas is traced via the $J$~=~1~$\rightarrow$~0 transition of $^{12}$CO, its luminosity implying total H$_{2}$ masses of (1.7~$\pm$ 0.3)~$\times$~10$^{11}$ and (9.5~$\pm$ 2.4)~$\times$~10$^{10}$~($\alpha_{\mathrm{CO}}$/0.8) M$_{\odot}$ in HATLAS\,J090426.9+015448 and HATLAS\,J140930.4+003803 respectively. Both galaxies exhibit molecular line emission over a broad ($\sim$1000 km/s) velocity range, and feature double-peaked profiles. We interpret this as evidence of either a large rotating disk or an on-going merger. Gas depletion timescales are $\sim$100 Myr. The 1.4~GHz radio luminosities of our targets place them close to the break in the luminosity function. As such they represent `typical' $z$~$>$~2 radio sources, responsible for the bulk of the energy emitted at radio wavelengths from accretion-powered sources at high redshift, and yet they rank amongst the most massive systems in terms of molecular gas and dust content. We also detect 115~GHz rest-frame continuum emission, indicating a very steep high-radio-frequency spectrum, possibly classifying the targets as compact steep spectrum objects.

\end{abstract}

\begin{keywords}
galaxies: high-redshift -- radio lines: galaxies -- radio continuum: galaxies
\end{keywords}




\section{Introduction}

Radio galaxies are one of several classes of object that harbour an active galactic nucleus \citep[AGN;][]{krawczynski13}. They are characterised by their extended radio emission which due to its high luminosity allows such objects to be detected out to very large distances. This property has allowed them to be used as cosmological probes since the discovery over half a century ago that the brightest radio sources in the sky were associated with galaxies at cosmologically significant distances \citep{minkowski60}. Identifying high redshift radio galaxies (HzRGs) is an effective way to pinpoint the sites of the most massive galaxies undergoing formation in the early Universe \citep{eales97,jarvis01,miley08}, as well as cluster environments at various stages of evolution \citep{stevens03,wylezelak13,hatch14,cooke16}. The present-day end-points of these systems are thought to be the giant elliptical galaxies that reside at the centres of rich clusters \citep{mclure99,orsi16}.

One of the most effective ways to study galaxies and galaxy groups at high redshift is to examine the cold gas in and around them via observations of molecular or atomic fine structure lines. A thorough review of this subject is given by \citet{carilli13}. The molecular phase of the interstellar medium (ISM) is dominated by H$_{2}$, however the usual observational tracers are the $J$~$\rightarrow$~($J$~$-$~1) rotational transitions of $^{12}$C$^{16}$O (hereafter CO), as this is the next most abundant molecule. The high excitation temperature of H$_{2}$ means it can only be observed directly in e.g.~the presence of strong shocks, however the low excitation temperature of 5.5~K that the CO molecule has means its ground-state transition can be used to trace even the normal, cold ISM within a galaxy. The 115.27 GHz rest frequency means that at the peak cosmological epoch of galaxy assembly, between redshifts of 1 and 3, the line is observable in the frequency ranges of many ground-based cm- and mm-wave facilities. 

Spatially and spectrally resolving this line allows much to be inferred about the dynamics of high redshift systems \citep[e.g.][]{hodge12}. The star forming potential of a galaxy can be established as the cold gas represents the fuel reservoir for this process, which when coupled with a star formation rate estimate allows gas consumption timescales to be determined \citep[e.g.][]{daddi10}. Thermodynamic modelling of the spectral energy distribution (SED) described by multiple $J$ transitions allows physical conditions of the ISM to be constrained \citep{obreschkow09}.

There has been a steady stream of studies of individual high-redshift systems via their CO lines since the first detection of the line at high redshift nearly a quarter of a century ago \citep[IRAS F10214+4724 at $z$~=~2.3;][]{brown91,solomon92}. Such studies have now looked back in cosmic time to the very first galaxies to form in the Universe \citep{wang13}. The last half-decade has seen a significant increase in the quantity and quality of the data being recorded in this field, largely due to significant upgrades to the bandwidth and sensitivity of existing radio interferometers such as the Karl G.~Jansky Very Large Array, the Plateau de Bure Interferometer (PdBI) and the Compact Array Broadband Back-end \citep[CABB;][]{wilson11} for the Australia Telescope Compact Array (ATCA), and the deployment of the Atacama Large Millimetre/submillimetre Array (ALMA). Broad bandwidths have unlocked the potential for blind molecular line scans for the first time, both for single objects \citep{harris12,vieira13} and for searches in extragalactic deep fields over significant cosmological volumes \citep{walter14,lentati14}. 

Radio galaxies being signposts for massive systems at high-$z$ makes them prime targets for follow-up molecular line observations, the first tentative CO detection in a HzRG being reported nearly twenty years ago by \citet{scoville97}. Since then numerous studies have shown that radio galaxies have molecular gas properties similar to those of quasars and, like quasars, often exhibit similarities to sub-mm galaxies (SMGs), e.g.~\citet{papadopoulos00}. Typical inferred H$_{2}$ masses exceed 10$^{10}$ M$_{\odot}$. Multiple components are often seen, corroborating the scenario that radio galaxy triggering is associated with merger activity or protocluster formation \citep{greve04,debreuck05,ivison12,ivison13}, a picture further supported by the often-disturbed optical morphologies seen at lower redshifts \citep{ramos11}. Several studies have reported alignments between radio jets and the peaks of the CO emission \citep{klamer05,emonts14}. The latter study reports CO peaks beyond the boundary of the radio emission, suggesting the radio jets play a role in stimulating star formation and metal enrichment. Molecular line emission has also been detected in the outer regions of radio galaxies \citep{nesvadba09}, likely associated with the Lyman-$\alpha$ haloes that extend for 10s--100s of kpc beyond the central radio galaxy \citep{jarvis03,wilman04}. A spectacular example of this is the `Dragonfly' galaxy at $z$~=~2, where 60\% of the total CO content is associated with tidal features within the halo \citep{emonts15}.

Here we add to the existing body of work on characterising the molecular gas in high redshift systems, presenting ATCA observations of CO~($J$~=~1~$\rightarrow$~0) in a pair of HzRGs. The 1.4 GHz radio luminosities place the two targets close to the break in the radio luminosity function, distinguishing them from existing molecular line studies that have typically targeted more powerful HzRGs \citep{klamer05,emonts11}. Our targets therefore represent the `typical' radio source at $z$~$>$~2, responsible for the bulk of the energy density emanating at radio wavelengths from high accretion rate radio galaxies. Throughout the paper the assumed cosmological parameters are: $H_{0}$~=~67.74 $\pm$ 0.46 km~s$^{-1}$~Mpc$^{-1}$, $\Omega_{\mathrm{M}}$~=~0.3089 $\pm$ 0.0062 and $\Omega_{\Lambda}$~=~0.6911 $\pm$ 0.0062 \citep{ade14}.
 
\section{Targets}
\label{sec:targets}

Our targets were selected from the sample of HzRGs studied by \citet{virdee13}, who investigated the radio properties of sources selected from the \emph{Herschel}-ATLAS \citep[H-ATLAS;][]{eales10} / Galaxy And Mass Assembly \citep[GAMA;][]{driver11} fields. A sample of seven radio sources were selected for follow-up observations, including long-slit spectroscopy with the Intermediate-dispersion Spectrograph and Imaging System (ISIS) on the William Herschel Telescope in order to determine redshifts, and Very Long Baseline Interferometry (VLBI) observations with the European VLBI Network (EVN) in order to disentangle the core, jet and star-formation components of the radio emission. The sample was constructed based on the fact that the far-infrared (FIR) and radio emission suggested that the sources harboured both an AGN, exhibited a significant amount of star-formation and were at $z$~$>$~1. The systems selected are also optically very faint. The spectroscopic follow-up by \citet{virdee13} revealed emission line properties that classified the target sources as narrow-line radio galaxies. Candidate lines were fitted to the spectral emission peaks and secondary lines were then searched for at their expected positions in order to confirm the candidate line. Our two targets only had single lines in each spectrum, which were assumed to be the bright Lyman-$\alpha$ line, on the basis that other typically bright lines (e.g.~C~IV, O~II) should have had secondary lines visible in the spectrum.

Far-infrared luminosities measured from the \emph{Herschel}-ATLAS data imply star-formation rates (SFRs) of $\sim$1000~M$_{\odot}$ / yr and high levels of dust (Section \ref{sec:lco}). 

We targeted two of these sources for ATCA follow-up to search for CO line emission as their redshifts placed the ground-state transition of CO in the 7~mm band of the ATCA. The properties of these two sources are given in Table \ref{tab:targets}, including the results derived in subsequent sections of this paper, as noted in the final column. 

\begin{table*}
\begin{minipage}{180mm}
\centering
\caption{Properties of HATLAS\,J090426.9+015448 and HATLAS\,J140930.4+003803. Parameters from existing observations are listed, together with the results derived in the subsequent sections of this paper, as listed in the final column. For the CO lines, each quantity is listed for the two components present in each spectrum, as labeled on Figures \ref{fig:JV09} and \ref{fig:JV15}, as well as for the system as a whole. It is not possible to conclusively differentiate between a major merger and a single massive molecular gas disk scenario for the two targets with the data in hand. The individual component measurements and the total measurements are listed in order to capture both possibilities, although we note that existing high-resolution observations of similar systems suggest that a merger is the most likely of the two \citep{genzel03,ivison10}.
\label{tab:targets}}
\begin{tabular}{llccccl} \hline
Quantity      & Units / Epoch  & \multicolumn{2}{c}{{\bf HATLAS\,J090426.9+015448}}            & \multicolumn{2}{c}{{\bf HATLAS\,J140930.4+003803}} & Reference \\ \hline
RA       & J2000        & \multicolumn{2}{c}{09h04m26.80s} & \multicolumn{2}{c}{14h09m30.23s} & \\ 
Dec      & J2000        & \multicolumn{2}{c}{+01\degr54\arcmin48.73\arcsec} & \multicolumn{2}{c}{+00\degr38\arcmin04.54\arcsec} & \\ 
$z_{\mathrm{Ly\alpha}}$ &  & \multicolumn{2}{c}{2.372} & \multicolumn{2}{c}{2.044} & \citet{virdee13}\\ 
$S_{\mathrm{100}}$ & mJy & \multicolumn{2}{c}{2.7 $\pm$ 46.8} & \multicolumn{2}{c}{113.1 $\pm$ 52.4} & H-ATLAS\\ 
$S_{\mathrm{160}}$ & mJy & \multicolumn{2}{c}{-57.7 $\pm$ 66.8} & \multicolumn{2}{c}{-50.07 $\pm$ 70.1} & H-ATLAS\\ 
$S_{\mathrm{250}}$ & mJy & \multicolumn{2}{c}{45.8 $\pm$ 7.0} & \multicolumn{2}{c}{38.5 $\pm$ 6.8} & H-ATLAS \\ 
$S_{\mathrm{350}}$ & mJy & \multicolumn{2}{c}{51.1 $\pm$ 8.0} & \multicolumn{2}{c}{47.1 $\pm$ 8.1} & H-ATLAS\\ 
$S_{\mathrm{500}}$ & mJy & \multicolumn{2}{c}{45.1 $\pm$ 8.8} & \multicolumn{2}{c}{22.0 $\pm$ 8.9} & H-ATLAS\\ 
$S_{\mathrm{500}}$ & mJy & \multicolumn{2}{c}{45.1 $\pm$ 8.8} & \multicolumn{2}{c}{22.0 $\pm$ 8.9} & H-ATLAS\\ 
$L_{\mathrm{1.4GHz}}$ & 10$^{26}$ W~Hz$^{-1}$~sr$^{-1}$ & \multicolumn{2}{c}{0.99 $\pm$ 0.07} & \multicolumn{2}{c}{6.25 $\pm$ 0.14} & Section~\ref{sec:contim}\\ 
$L'_{\mathrm{CO}}$      & K km s$^{-1}$ pc$^{2}$                   & \multicolumn{2}{c}{(2.11 $\pm$ 0.37) $\times$ 10$^{11}$} & \multicolumn{2}{c}{(1.18 $\pm$ 0.30) $\times$ 10$^{11}$} & Section~\ref{sec:lco}\\ 
$M_{\mathrm{H_{2}}}$    & ($\alpha_{\mathrm{CO}}$/0.8)~M$_{\odot}$ & \multicolumn{2}{c}{(1.69 $\pm$ 0.29) $\times$ 10$^{11}$} & \multicolumn{2}{c}{(9.45 $\pm$ 2.36) $\times$ 10$^{10}$} & Section~\ref{sec:lco}\\ 
$L_{\mathrm{FIR}}$ (8--1000~$\mu$m)     & L$_{\odot}$                              & \multicolumn{2}{c}{(11.75 $\pm$ 1.62) $\times$ 10$^{12}$} & \multicolumn{2}{c}{(7.59 $\pm$ 1.22) $\times$ 10$^{12}$}& Section~\ref{sec:lco}\\ 
$q_{\mathrm{IR}}$ &      & \multicolumn{2}{c}{1.92~$\pm$~0.13} & \multicolumn{2}{c}{0.81~$\pm$~0.10}& Section~\ref{sec:lco}\\ 
SFR                     & M$_{\odot}$~yr$^{-1}$                    & \multicolumn{2}{c}{1250~$\pm$~160} & \multicolumn{2}{c}{810~$\pm$~130} & Section~\ref{sec:lco}\\
$T_{\mathrm{dust}}$     & K                             & \multicolumn{2}{c}{30~$\pm$~4} & \multicolumn{2}{c}{31~$\pm$~6}& Section~\ref{sec:lco}\\ 
$L_{\mathrm{dust}}$     & L$_{\odot}$                              & \multicolumn{2}{c}{(11.5~$\pm$~1.6)~$\times$~10$^{12}$} & \multicolumn{2}{c}{(7.9~$\pm$~1.1)~$\times$~10$^{12}$}& Section~\ref{sec:lco}\\ 
$M_{\mathrm{dust}}$     & M$_{\odot}$                              & \multicolumn{2}{c}{(2.3 $\pm$ 1.1) $\times$ 10$^{9}$} & \multicolumn{2}{c}{(1.3 $\pm$ 0.7) $\times$ 10$^{9}$} & Section~\ref{sec:lco}\\ 
$M_{\mathrm{dyn}}$ sin$^{2}$($i$)& M$_{\odot}$                     & \multicolumn{2}{c}{$<$(5.67~$\pm$~0.17)~$\times$~10$^{12}$} & \multicolumn{2}{c}{$<$(9.24~$\pm$~0.27)~$\times$~10$^{11}$}& Section~\ref{sec:velocity}\\ \hline
           &            & \multicolumn{1}{c}{{\bf Component 1}} & \multicolumn{1}{c}{{\bf Component 2}}    & \multicolumn{1}{c}{{\bf Component 1}}  & \multicolumn{1}{c}{{\bf Component 2}}& Figures \ref{fig:JV09},\ref{fig:JV15}\\ \hline
$\nu_{\mathrm{centre}}$ & GHz                    & 34.1475 $\pm$ 0.0040   & 34.2397 $\pm$ 0.0022    & 37.8293 $\pm$ 0.0038             & 37.8708 $\pm$ 0.0056     & Section~\ref{sec:spectra}\\ 
$S_{\mathrm{peak}}$     & mJy beam$^{-1}$        & 0.352 $\pm$ 0.145      & 0.5821 $\pm$ 0.0575     & 0.436 $\pm$ 0.077                & 0.374 $\pm$ 0.061        & Section~\ref{sec:spectra}\\ 
$v_{\mathrm{FWHM}}$     & km s$^{-1}$            & 173 $\pm$ 80           & 414 $\pm$ 46            & 228 $\pm$ 63                     & 318 $\pm$ 103            & Section~\ref{sec:spectra}\\
$z_{\mathrm{CO}}$       &                        & 2.3756 $\pm$ 0.0004    & 2.3666 $\pm$ 0.0002     & 2.0471 $\pm$ 0.0003              & 2.0437 $\pm$ 0.0004      & Section~\ref{sec:spectra}\\
$D_{\mathrm{L}}$        & Gpc                    & 19.652 $\pm$ 0.293     & 19.562 $\pm$ 0.291      & 16.397 $\pm$ 0.242               & 16.364 $\pm$ 0.242       & Section~\ref{sec:spectra}\\
$I_{\mathrm{CO}}$       & Jy km s$^{-1}$         & 0.153 $\pm$ 0.095      & 0.605 $\pm$ 0.090       & 0.249 $\pm$ 0.082                & 0.299 $\pm$ 0.109        & Section~\ref{sec:lco}\\ 
$L'_{\mathrm{CO}}$      & K km s$^{-1}$ pc$^{2}$ & (4.27 $\pm$ 2.66) $\times$ 10$^{10}$ & (1.68 $\pm$ 0.25) $\times$ 10$^{11}$   & (5.37 $\pm$ 1.78) $\times$ 10$^{10}$    & (6.43 $\pm$ 2.35) $\times$ 10$^{10}$ & Section~\ref{sec:lco}\\ 
$M_{\mathrm{H_{2}}}$ & ($\alpha_{\mathrm{CO}}$/0.8)~M$_{\odot}$ & (3.42 $\pm$ 2.13) $\times$ 10$^{10}$ & (1.35 $\pm$ 0.20) $\times$ 10$^{11}$ & (4.30 $\pm$ 1.32) $\times$ 10$^{10}$ & (5.15 $\pm$ 1.88) $\times$ 10$^{10}$ & Section~\ref{sec:lco}\\
$M_{\mathrm{dyn}}$ sin$^{2}$($i$) & M$_{\odot}$ & $<$(1.4~$\pm$~1.3)~$\times$~10$^{11}$ & $<$(8.1~$\pm$~1.8)~$\times$~10$^{11}$ & $<$(1.1~$\pm$~0.6)~$\times$~10$^{11}$&$<$(2.1~$\pm$~1.4)~$\times$~10$^{11}$ & Section~\ref{sec:velocity}\\ \hline
\end{tabular}
\end{minipage}
\end{table*}

\section{Observations and data reduction}

We made use of the ATCA in its most compact H75 configuration, the northern spur of the array being necessary due to the proximity of our targets to the celestial equator. Observations\footnote{Project code: C2847} were conducted over five nights for each target, avoiding antenna elevations below 30 degrees in order to keep the system temperature low. The 7~mm receivers were used, with the CABB configured to deliver 2~$\times$~2 GHz basebands, each consisting of 2,048~$\times$~1~MHz channels. For this project we made use of only one of the basebands, namely those that were expected to contain the CO line. For HATLAS\,J090426.9+015448 the frequency coverage was 33.416--35.464~GHz, and for HATLAS\,J140930.4+003803 it was 37.100--39.148~GHz. 

The strong source PKS 1253$-$055 (3C 279) was observed once per night in order to calibrate the bandpass, and we set the flux density scale using the standard calibrator PKS B1934-638. Nearby ($<$2 degrees separation), compact secondary calibrator sources PKS B0906+015 for HATLAS\,J090426.9+015448, and PKS B1356+022 for HATLAS\,J140930.4+003803) were observed for two minutes after every ten minutes of target observation in order to calibrate the complex antenna gains. 

\subsection{Calibration}

The ten nights' worth of data were initially loaded into the {\sc miriad} package \citep{sault95}, using {\sc atlod}, which was set to automatically flag channels that are known to contain radio frequency interference (RFI) as well as the band edges where the gain drops due to the filter responses. Autocorrelations were also discarded at this stage. Antenna positions were updated using {\sc atfix}. Following this step, the data were converted to Measurement Set format using {\sc casa}'s (McMullin et al., 2007) {\sc importmiriad} task, and the baseband relevant to the CO emission was extracted using the {\sc split} task. The ten Measurement Sets were examined using {\sc plotms} and any obviously bad data were removed. Antenna 6 was also discarded outright. It forms baselines with the other five antennas that are approximately 4.4~km in length, the next longest baseline for the H75 configuration being 89~m. Such a large gap in the ($u$,$v$) plane coverage does not usefully contribute to a higher resolution image without discarding most of the sensitivity provided by the inner baselines. The {\sc bandpass} task was used to derive normalised, time-independent solutions for each 1~MHz frequency channel from the scans of 1253$-$055. A single frequency-independent complex gain correction was derived for each scan of the secondary calibrator using the {\sc gaincal} task. The flux density scale of these solutions were then corrected using the {\sc fluxscale} task and a model for PKS B1934-638. The bandpass and flux-calibrated complex gain solutions were then applied to the target sources, using a linear interpolation in time for the latter set of corrections, via the {\sc applycal} task. Note that the {\sc casa setjy} model for PKS B1934$-$638 does not reliably extend to these frequencies. The polynomial spectral model provided by the ATCA Calibrator Database\footnote{{\tt \href{http://www.narrabri.atnf.csiro.au/calibrators/}{http://www.narrabri.atnf.csiro.au/calibrators/}}} was implemented instead. The {\sc setjy} task was mimicked by producing per-channel point source models of appropriate brightness using {{\sc casa}'s {\sc cl} tool, and filling the {\tt MODEL} column of the Measurement Set using the {\sc ft} task. The total on-source observing time following the removal of bad data was 16.13 and 22.28 hours for HATLAS\,J090426.9+015448 and HATLAS\,J140930.4+003803 respectively.

\subsection{Imaging and continuum subtraction}
\label{sec:imaging}

Following calibration, a full-band continuum image of each target was made using Multi-Frequency Synthesis (MFS) imaging. No deconvolution was applied initially. The dirty map was inspected and revealed a compact feature at the expected position of each target. This step was necessary because the declinations of our targets result in point spread functions (PSFs, or dirty / synthesised beams) with pronounced north-south sidelobe features that rise to up to 95 percent of the peak value. Any deconvolution must therefore be done with extreme caution to avoid sidelobes being interpreted as genuine emission and biasing the true flux density measurement. Having identified the central peak and compared the associated sidelobes to those of the PSF, a shallow (100 clean iterations) deconvolution was allowed to proceed, and within a region that contained only the central peak. The resulting images are presented in Section \ref{sec:contim}. The restoring beams used in the continuum images are 2D Gaussians with full-width at half-maximum (FWHM; major axis $\times$ minor axis) extents of 19$\arcsec$~$\times$~15$\arcsec$ (position angle~=~$-$57$\degr$, east of north) and 16$\arcsec$~$\times$~14$\arcsec$ (position angle~=~$-$82$\degr$) for HATLAS\,J090426.9+015448 and HATLAS\,J140930.4+003803 respectively.

In order to search for spectral line emission the continuum emission was subtracted using the {\sc casa}'s {\sc uvcontsub} task. This task estimates the continuum emission by fitting polynomials (in this case a simple first-order polynomial) to the real and imaginary parts of the visibility spectrum \citep[e.g.][]{sault94}. A solution was generated for each ten-minute scan of the target, and the resulting model was subtracted from the visibilities. Following this, image cubes were made, with no deconvolution, and spectra were extracted at the peak position of the continuum emission. Incremental averaging in frequency was used to boost the signal-to-noise ratio of the line emission. A spectral averaging factor of 6 was found to reveal a spectrum that had both high significance and adequate velocity resolution. Having identified the region of the spectrum over which line emission was present, the continuum subtraction was repeated, excluding from the fit the regions where the line emission was known to be. This was mainly done out of caution, and made no discernible difference, as one would expect from a low-order polynomial fit and a line width that is narrow compared to the total bandwidth. The resulting spectra are presented in Section \ref{sec:spectra}. The velocity resolutions of each 6~MHz channel are 53 and 46 km s$^{-1}$ for HATLAS\,J090426.9+015448 and HATLAS\,J140930.4+003803 respectively, however note that Hanning smoothing was applied to the plotted spectra to emphasise the line emission, which doubles the effective velocity resolution of each channel in the figures.

\section{Results and discussion}

\begin{figure*}
\begin{center}
\setlength{\unitlength}{1cm}
\includegraphics[width = \textwidth]{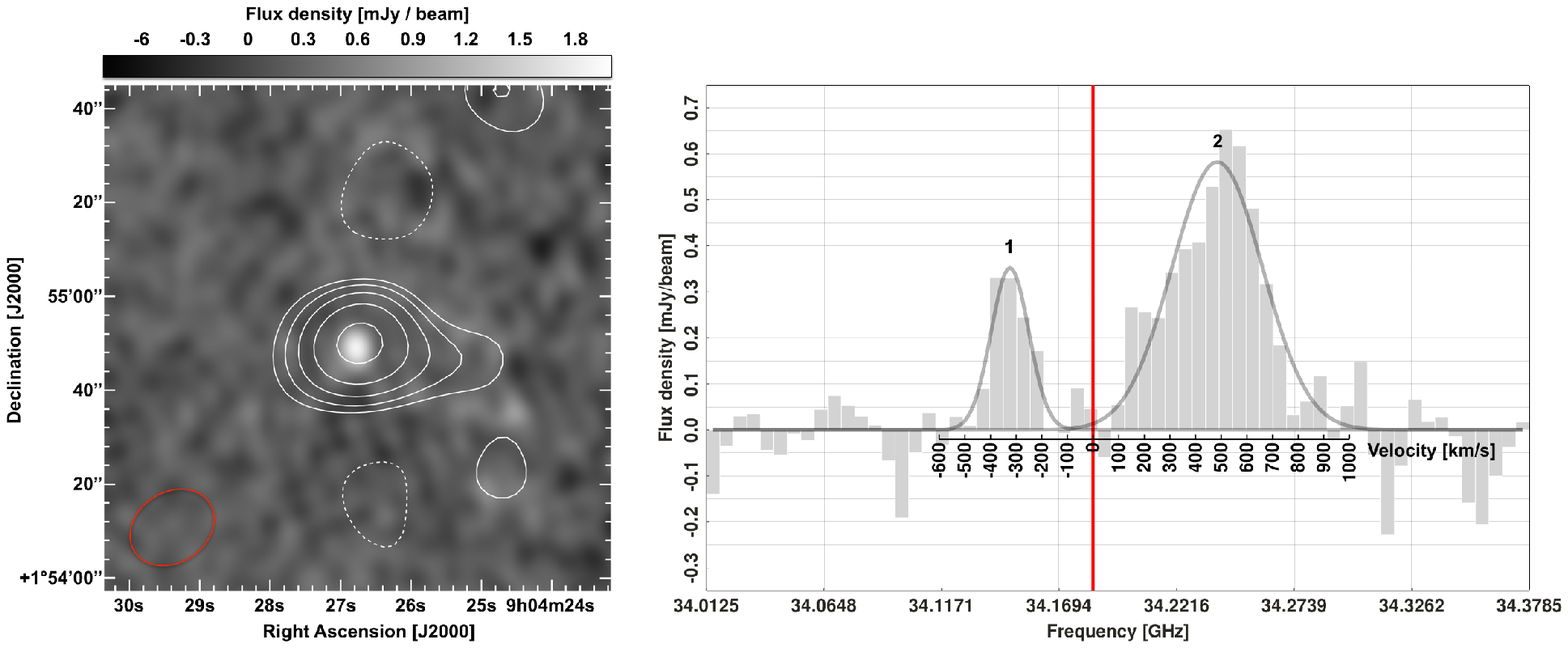}
\caption{HATLAS\,J090426.9+015448 ($z$~=~2.372, marked with vertical line). {\bf Left panel:} The ATCA-detected radio continuum emission from HATLAS\,J090426.9+015448. The greyscale shows the 1.4~GHz FIRST detection and the contours show the ATCA-detected radio emission (SNR~=~14) at 34.44~GHz (115~GHz rest-frame). Contour levels are 22~$\times$~(1, $\sqrt{2}$, 2, 2$\sqrt{2}$, 4) $\mu$Jy / beam. There is a single negative contour at $-$22 $\mu$Jy / beam. The 1$\sigma$ background noise level in the ATCA map is 14 $\mu$Jy / beam. The extended feature to the west of the peak is not believed to be significant, a result of residual phase errors. The restoring beam used following deconvolution is a Gaussian with a major axis of 19 arcseconds, a minor axis of 15 arcseconds and a position angle of $-$57 degrees east of north, as indicated by the ellipse in the lower left. {\bf Right panel:} Histogram showing the CO ($J$~=~1~$\rightarrow$~0) spectrum of HATLAS\,J090426.9+015448. The vertical line is the Lyman-$\alpha$ redshift \citep{virdee13} and the velocity of the molecular gas with respect to this is shown on the panel. The solid lines show the two Gaussian components fitted to the peaks as described in Section \ref{sec:lco}. The 1$\sigma$ image noise values, measured in each plane of the image cube over the channels shown above, has a mean value of 0.21 mJy / beam with a standard deviation of 0.02 mJy / beam. The noise in the plotted spectrum appears considerably lower due to the effects of the Hanning smoothing.
\label{fig:JV09}}  
\end{center}
\begin{center}
\setlength{\unitlength}{1cm}
\includegraphics[width = \textwidth]{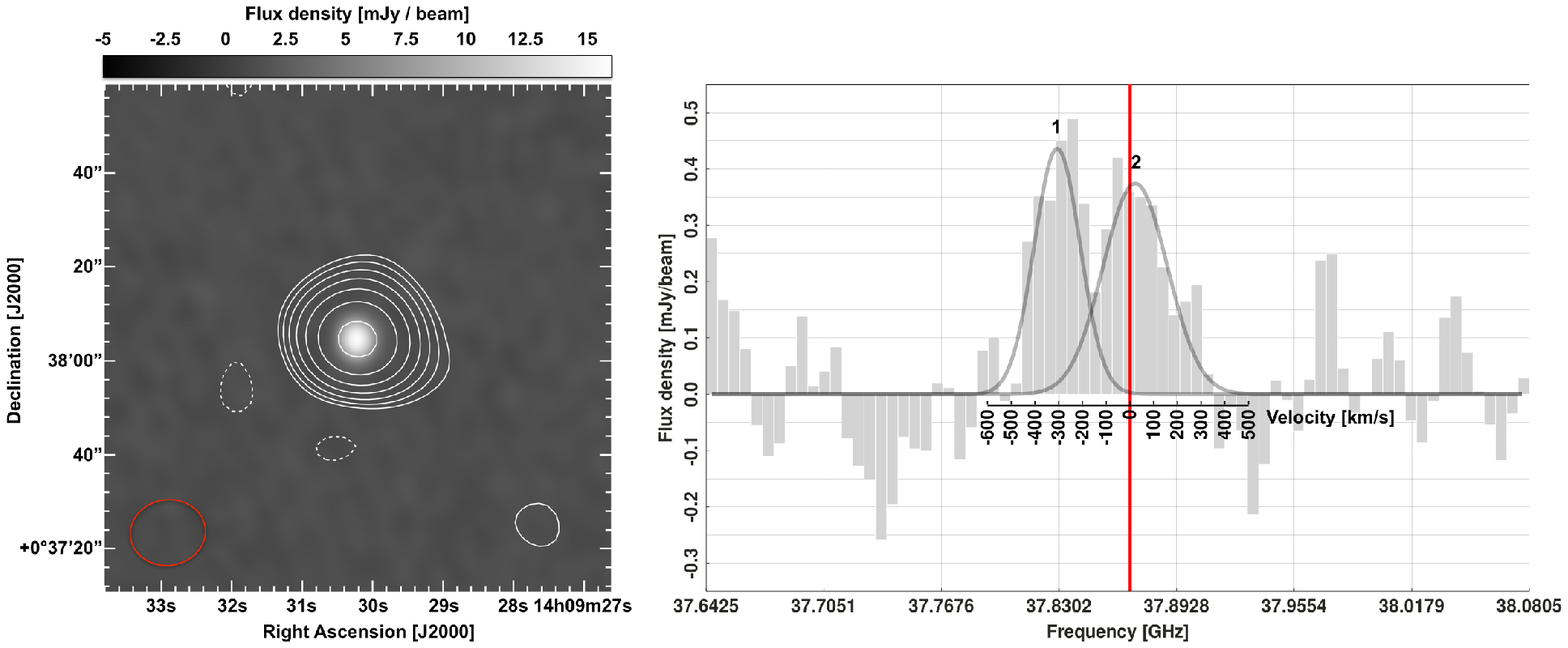}
\caption{HATLAS\,J140930.4+003803 ($z$~=~2.044, marked with vertical line). {\bf Left panel:} The ATCA-detected radio continuum emission from HATLAS\,J140930.4+003803. Greyscale and contours are as per Figure \ref{fig:JV09} above. The 38.124~GHz (115~GHz rest-frame) ATCA detection has an SNR of 18. Contour levels are 20~$\times$~(1, $\sqrt{2}$, 2, 2$\sqrt{2}$, 4, 4$\sqrt{2}$, 8) $\mu$Jy / beam, with a single negative contour at $-$16 $\mu$Jy / beam. The 1$\sigma$ noise level in the ATCA map is 16 $\mu$Jy / beam. The restoring beam has a major axis of 16 arcseconds, a minor axis of 14 arcseconds and a position angle of $-$82 degrees east of north, as shown in the lower left. {\bf Right panel:} Histogram showing the CO ($J$~=~1~$\rightarrow$~0) spectrum of HATLAS\,J140930.4+003803. The vertical line is the Lyman-$\alpha$ redshift \citep{virdee13}. Properties of the two Gaussian components fitted to the peaks are described in Section \ref{sec:lco}.  The 1$\sigma$ image noise values, measured in each plane of the image cube over the channels shown above, has a mean value of 0.22 mJy / beam with a standard deviation of 0.02 mJy / beam. Again, the Hanning smoothing makes the noise in the plotted spectrum appear much lower.
\label{fig:JV15}}  
\end{center}
\end{figure*}

\subsection{Continuum detections at 115 GHz rest-frame}
\label{sec:contim}

Both of our targets are detected in the ATCA continuum maps, presented in the left-hand panels of Figures \ref{fig:JV09} and \ref{fig:JV15}. These images show the ATCA contours (115~GHz rest-frame) overlaid on the corresponding 1.4 GHz images from the Faint Images of the Radio Sky at Twenty-cm \citep[FIRST;][]{becker95} survey. Contour and greyscale levels are provided in the figure captions. Component fitting to the ATCA detections was performed using the {\sc casa} task {\sc imfit}, which determined the emission to be point-like for both sources. The fitted position for HATLAS\,J090426.9+015448 is at right ascension 09h04m26.711s $\pm$ 0.068~s and declination +01\degr54\arcmin49.67\arcsec $\pm$~0.58\arcsec, with a peak flux density of 160~$\pm$~16~$\mu$Jy beam$^{-1}$. For HATLAS\,J140930.4+003803 the fitted right ascension and declination are 14h09m30.211s $\pm$~0.020~s and +00\degr38\arcmin04.62\arcsec $\pm$ 0.23\arcsec, with a peak flux density of 310 $\pm$ 12 $\mu$Jy beam$^{-1}$. Note that the quoted positional errors are those derived from the statistical uncertainties in the component fitting procedure and do not contain any astrometric reference frame errors, however the latter should be negligible. No self-calibration has taken place so the astrometric accuracy due to the calibration will be tied to the position of the phase calibrator.

Radio continuum spectra for both of the targets are shown in Figure \ref{fig:conspec}, which features both the NVSS measurement as well as the 327~MHz Giant Metrewave Radio Telescope (GMRT) detection from the catalogues of \citet{mauch13}. 
The 1.4~GHz radio luminosities (L$_{\mathrm{1.4GHz}}$) in Table \ref{tab:targets} are derived from the NRAO VLA Sky Survey \citep[NVSS;][]{condon98} flux density measurements and the mean luminosity distances listed in Table \ref{tab:targets}.

The spectra for both objects exhibit a break at high frequencies. While the spectral break may be explained in terms of the higher frequency emission being core-dominated and the lower frequency emission containing a mixture of core and extended synchrotron emission \citep[e.g.][]{whittam13,whittam16}, the spectra of these two objects are atypical for HzRGs. \citet{klamer06} conducted a search for HzRGs based a spectral steepness selection, finding that the 89\% of their sample have radio spectra that are well-described by a simple power law, with only 11\% exhibiting mild spectral curvature. The two HzRGs studied via their CO lines by \citet{emonts11} also display no spectral curvature between 1.4 and 30~GHz (observed frame). \citet{herzog16} also report that a high fraction ($\sim$75\%) of the radio SEDs of infrared-faint radio sources (IFRS) can be modelled by a simple power law, extending as far as 105~GHz (observed frame). The radio properties of IFRS are reported to be consistent with the general population of radio-loud AGN at high redshift, with similar fractions of compact steep spectrum (CSS) and gigahertz-peaked spectrum (GPS) sources. CSS and GPS sources have concave radio spectra with steep spectral slopes either side of a peak \citep[see][for a review]{odea98}. Any extended radio emission associated with these sources exists in a small region, well within the host galaxy, either due to youth or environmental confinement \citep[e.g.][]{marr14}. If the spectral peak for our two targets, lying somewhere below 1~GHz, gives way to a low-frequency turnover they could reasonably be classified as faint, high-redshift CSS sources.

The 1.4--35 GHz (observed frame) spectral indices ($\alpha$, where flux density $S$~$\propto$~$\nu^{\alpha}$) as measured from Figure \ref{fig:conspec} are $-$3.1 $\pm$ 0.1 and $-$3.34 $\pm$ 0.03 for HATLAS\,J090426.9+015448 and HATLAS\,J140930.4+003803 respectively. Even for a steep-spectrum radio galaxy \citep[e.g.][]{sadler14} this is excessively steep. Returning to our previous possible explanation for the observed spectral break, the continuum spectrum may be rendered artificially steep if the existing lower frequency measurements are capturing both core and extended emission, and our ATCA observations are sensitive only to a flatter-spectrum core contribution. Inverse-Compton losses due to the scattering of electrons by Cosmic Microwave Background photons may also be important at this redshift \citep{murphy09} as a mechanism for suppressing the synchrotron emission. In the event that any extended jet-driven emission is compact (as it would be for CSS or GPS sources) an artificial steepening of the measured spectrum would be more likely occur for multifrequency radio observations with (approximately) matched angular resolution that is insufficient to separate core and jet structures, as we have used here. It is noteworthy that there is no discrepancy between the FIRST and NVSS measurements for HATLAS\,J140930.4+003803, however for HATLAS\,J090426.9+015448 the peak flux density measurement from FIRST is 28\% lower than the NVSS measurement, suggesting possible extended emission on scales greater than those corresponding to the synthesised beam of FIRST (50~kpc at $z$~=~2.37). The EVN imaging presented by \citet{virdee13} resulted in no significant detection of emission associated with the latter. For the former there was a tentative detection of two components, with a separation of 0.8\arcsec (6.8~kpc projected on the plane of the sky at $z$~=~2.04) and a combined integrated flux density of $\sim$12~mJy, consistent with the CSS classifcation.

With the data-in hand the only solid conclusion that we may draw is that the radio spectra of these two objects are intriguing. Further observations at intermediate frequencies, as well as sub-327~MHz would be needed to accurately constrain the shape and peak of the continuum spectra for these two sources, and with angular resolution sufficient to separate any extended structures from core emission. Observations of the GAMA fields with the Low Frequency Array (LOFAR) are pending as part of its campaign to observe the \emph{Herschel}-ATLAS fields \citep{hardcastle16}.

\begin{figure}
\centering
\includegraphics[width=\columnwidth]{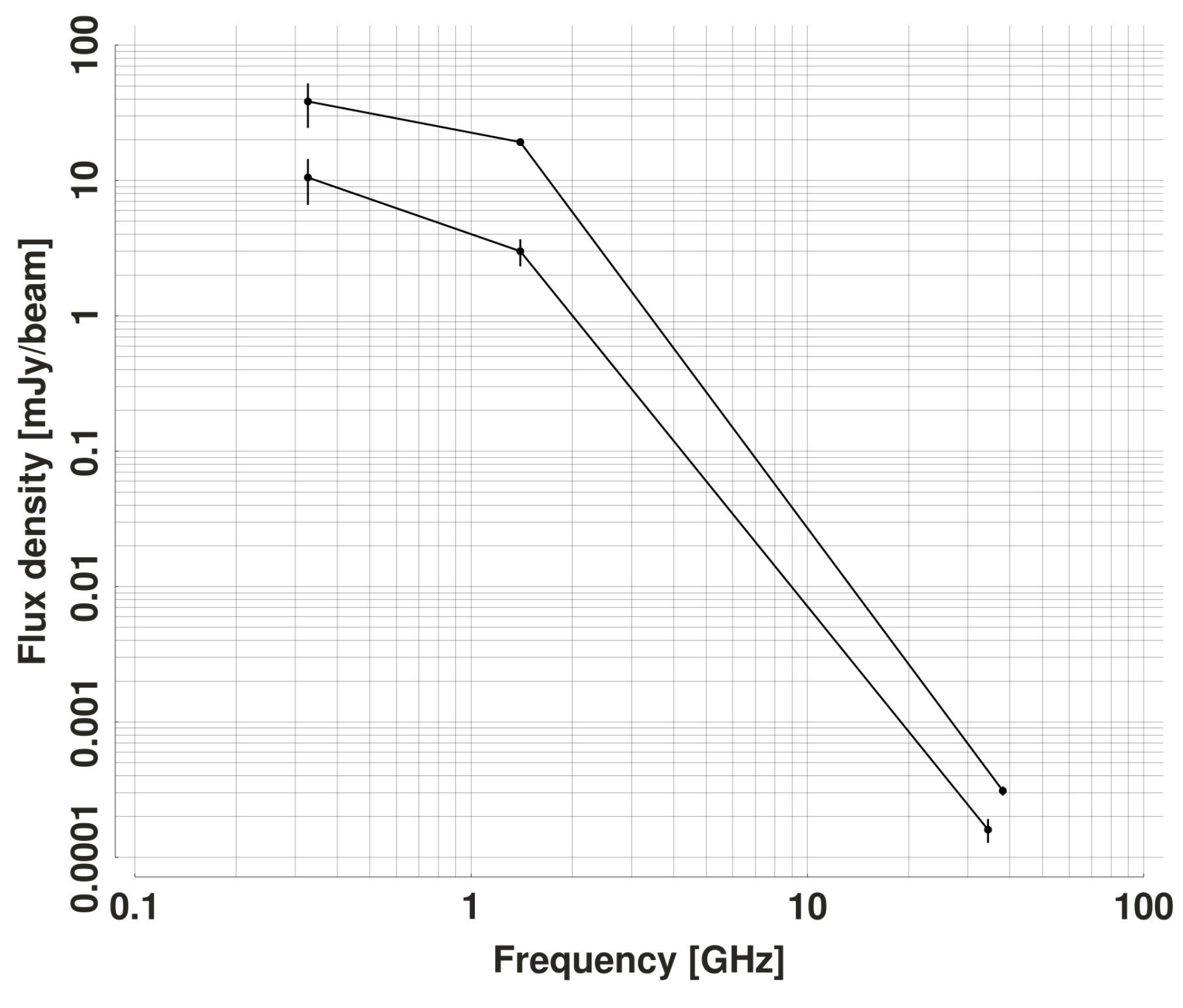}
\caption{Radio continuum spectra of HATLAS\,J090426.9+015448 and HATLAS\,J140930.4+003803. In increasing frequency order the measurements are from the 325~MHz GMRT survey of \citet{mauch13}, the 1.4 GHz NVSS survey \citep{condon98} and this work.}
\label{fig:conspec}
\end{figure}

\subsection{CO~($J$~=~1~$\rightarrow$~0) detections}
\label{sec:spectra}

The spectra of HATLAS\,J090426.9+015448 and HATLAS\,J140930.4+003803 extracted from the continuum-subtracted data at the pixel coincident with the peak of the continuum emission are shown in the right hand panels of Figures \ref{fig:JV09} and \ref{fig:JV15} respectively. There is strong line emission present, confirming the redshift of what was assumed to be the Lyman-$\alpha$ line as measured by \citet{virdee13}, at the expected frequency for CO ($J$~=~1~$\rightarrow$~0). Note that \citet{virdee13} provides no estimate of the uncertainty in the Lyman-$\alpha$ redshift. 

Velocity-integrated (moment zero) maps formed from the continuum-subtracted channels containing the line emission did not reveal any plausible extended emission. This is perhaps unsurprising given the low angular resolution of the observations, however without the benefits afforded by MFS, the deconvolution of the narrow channel cubes was not straightforward. Many sidelobe-like features remained, possibly exacerbated by residual, low-level phase calibration errors. The Lyman-$\alpha$ redshift is marked on each spectrum with a vertical line, and the velocity of the molecular gas with respect to this is also shown on the figure. Velocity resolution in each 6~MHz channel is 53 and 46 km s$^{-1}$ for HATLAS\,J090426.9+015448 and HATLAS\,J140930.4+003803 respectively, although the effective velocity resolutions of the plotted spectra are twice these values due to the application of Hanning smoothing. The CO lines are well modelled by pairs of Gaussians in flux-density / frequency space, and these fits (to the non-smoothed data) are also shown on the figure. The RMS of the pixel values is measured over the spatial dimensions of each 6~MHz channel in the data cube, and this corresponding quantity is attached to each value in the line spectrum as an estimate of the uncertainty. The mean and standard deviations of these error spectra are provided in the captions of Figures \ref{fig:JV09} and \ref{fig:JV15}. Note that with the application of Hanning smoothing to the \emph{plotted} spectra, the apparent noise properties will be deceptively low. Adjacent channels become correlated depending on the width of the filter function, resulting not only in a coarser effective channel resolution but a significantly broader noise equivalent bandwidth for each channel.

The best fitting parameters and their associated uncertainties for the four components are provided in Table \ref{tab:targets}, with the full-width at half-maximum (FWHM) frequency values ($\nu_{\mathrm{centre}}$) being converted to a velocity FWHM ($v_{\mathrm{FWHM}}$) about the component centre, and the central frequency also being converted to a redshift ($z_{\mathrm{CO}}$) for CO ($J$~=~1~$\rightarrow$~0). The total velocity-integrated line flux for each component ($I_{\mathrm{CO}}$) is also given. These parameters will feature again in the discussion in Sections \ref{sec:velocity} and \ref{sec:lco}.

\subsection{Velocity structure}
\label{sec:velocity}

Frequency structure in atomic and molecular lines that are not spatially resolved can still reveal much about the gas velocities in galaxies and galaxy groups, as well as provide dynamical mass limits. Probably the best studied examples are local extragalactic systems observed via their neutral hydrogen (HI) lines \citep[e.g.][]{koribalski04}. HI lines in disk-dominated systems exhibit characteristic inclination-dependent double-peaked profiles. The profiles tend towards single peaks as the inclination angle\footnote{We follow the convention that $i$~=~0\degr~for a face-on system.} ($i$) approaches 0\degr and the extent of the profile becomes dominated by the velocity dispersion internal to the disk, assuming the (typical) case that dispersion velocity is small compared to the rotation velocity. HI double-peaked profiles are often asymmetric, possible reasons for which include warped disks or otherwise asymmetric rotating gas distributions. Profiles consistent with disks are also seen in numerous high redshift CO observations, a set of prime examples of which are presented by \citet{daddi10}. Their CO~($J$~=~2~$\rightarrow$~1) observations with the PdBI at 92~GHz reveal significant disks of cold molecular gas in a sample of star-forming galaxies at $z$~$\sim$~1.5. The velocity profiles of the sample show single- and double-peaked structures, the latter exhibiting both symmetric and highly asymmetric profiles. Cases of highly irregular high redshift CO line profiles have also been observed \citep[e.g.][]{riechers08}, suggestive of highly disrupted systems with numerous components. Numerous studies present substantial evidence that on-going mergers are responsible for the observed CO line profiles of many systems \citep[e.g.][]{frayer08,engel10,ivison10}, including examples that see evidence of mergers between gas-rich disks \citep{ivison11,ivison13}. Gravitational lensing plays a part in many high redshift systems for which CO lines have been observed, and this can also skew the velocity profile due to differential lensing \citep{deane13,deane15}.

The line structure in HATLAS\,J090426.9+015448, shown in the right hand panel of Figure \ref{fig:JV09}, exhibits an asymmetric double-peaked profile that covers a total velocity width of approximately 1100 km s$^{-1}$. This is consistent with the broadest CO line widths observed in SMGs, e.g.~\citet{harris12}. The vertical line on Figure \ref{fig:JV09} is the redshift of the Lyman-$\alpha$ line measured by \citet{virdee13}, which lies at the centre of the two CO peaks. If the Lyman-$\alpha$ redshift is coincident with the central AGN then a plausible explanation for the line structure is that the radio galaxy HATLAS\,J090426.9+015448 is surrounded by a large, asymmetric disk of molecular gas. The line profile is similar to that of SMM J02399$-$0136 as reported by \citet{genzel03}, who interpreted it as a single, rapidly rotating disk. Lyman-$\alpha$ lines can however be offset from cold gas tracer lines by $\sim$100s of km~s$^{-1}$ \citep[e.g.][]{willott15}, and in a manner that is different from object to object due to the resonant absorption on the blue side of the Lyman-$\alpha$ line. The systemic redshift of the AGN could well be centred on one of the CO peaks, and the double profile could be the result of two separate galaxies in the process of merging. SMM J02399$-$0136 was indeed later revealed to contain multiple merging objects \citep{ivison10}. Higher resolution optical spectroscopy of the Lyman-$\alpha$ line, or preferably longer wavelength observations of non-resonant lines would be needed to more accurately measure the systemic AGN redshift.

Our second target HATLAS\,J140930.4+003803 also exhibits two distinct peaks in the CO line, however in this case the profile is much more symmetric. The centre of component 2 (as per the labelling on Figure \ref{fig:JV15}) is coincident with the measured Lyman-$\alpha$ redshift. Again if the Lyman-$\alpha$ line is coincident with the central AGN the line profile can be explained by an on-going merger, with component 2 representing a molecular gas reservoir centred on the AGN and component 1 representing a second system. A second plausible explanation is that the CO line profile is the result of a single, regular, rotating disk of molecular gas, and either the Lyman-$\alpha$ redshift is slightly offset from the systemic redshift of the AGN, or the AGN itself is offset from the centre of the disk.

We can place upper limits on the dynamical masses in both systems, for both the merger or single disk scenario. The dynamical mass of a rotating disk (M$_{\mathrm{dyn}}$) in solar masses can be estimated as	
\begin{equation}
M_{\mathrm{dyn}} \sin^{2}(i) = 4 \times 10^{4} \Delta v^{2}R
\label{eq:mdyn}
\end{equation}
where $i$ is the inclination angle, $\Delta v$ is the FWHM of the line width in km~s$^{-1}$ and $R$ is the outer radius of the disk in kpc \citep{neri03}. Attempts to deconvolve the PSF and estimate the angular extent of the sources result only in upper limits of 14$''$ and 6$''$ for HATLAS\,J090426.9+015448 and HATLAS\,J140930.4+003803 respectively. The assumed cosmological model and the redshifts of the sources result in angular diameter distances that translate to angular scales of 8.37~$\pm$~0.25 and 8.56~$\pm$~0.25 kpc per arcsecond for the two targets, and the product of these and the angular size upper limits give upper limits to the value of $R$. We note that the typical extents of similar high-$z$ sources observed in the ground-state of CO are $\sim$5--15 kpc \citep{ivison11,riechers11,hodge12}, approximately an order of magnitude smaller than the constraints we are able to place on our targets using the current data. Crude dynamical mass upper limits can be estimated for each pair of components in each source by using the fitted velocities listed in Table \ref{tab:targets}. For HATLAS\,J090426.9+015448 we obtain values of M$_{\mathrm{dyn}} \sin^{2}(i)$~$<$~(1.4~$\pm$~1.3)~$\times$~10$^{11}$ and $<$(8.1~$\pm$~1.8)~$\times$~10$^{11}$ M$_{\odot}$ for components 1 and 2 respectively. The corresponding values for components 1 and 2 in HATLAS\,J140930.4+003803 are $<$(1.1~$\pm$~0.6)~$\times$~10$^{11}$ and $<$(2.1~$\pm$~1.4)~$\times$~10$^{11}$ M$_{\odot}$. The total dynamical mass limits are $<$5.7~$\times$~10$^{12}$~M$_{\odot}$ and $<$9.2~$\times$~10$^{11}$~M$_{\odot}$, assuming velocity widths of 1100 and 665 km~s$^{-1}$ for HATLAS\,J090426.9+015448 and HATLAS\,J140930.4+003803 respectively.

These (very conservative) dynamical mass limits for the individual components are consistent with existing measurements in other systems. One of the best examples of a dynamical study of a high-$z$ CO disk is the SMG GN20 (J123711.89+622211.8). \citet{hodge12} comfortably resolve a clumpy 14~kpc molecular disk in this system and derive a dynamical mass of (5.4 $\pm$ 2.4)~$\times$~10$^{11}$ M$_{\odot}$. \citet{wang13} report values of M$_{\mathrm{dyn}} \sin^{2}(i)$ of order 10$^{10}$--10$^{11}$~M$_{\odot}$ in their sample of z~$\sim$~6 QSOs. SMGs tend to have broader linewidths and therefore higher dynamical mass estimates than QSOs, and as noted by \citet{coppin10} this may be expected from orientation arguments when targeting optically selected QSOs under the assumption of a unified AGN model. However there are overlaps and exceptions in both populations of sources. Invoking a unified model would imply that HzRGs should have a broader range of linewidths and dynamical masses, similar to the typical SMG population. Ultimately, it must be stressed that our dynamical mass limits are simply not that stringent due to the relatively coarse angular resolution afforded by the ATCA's H75 configuration.

The possibility that the CO spectra also represent an AGN-driven outflow should be considered. Velocities of $\sim$1000 km~s$^{-1}$ are seen in many systems and via many spectral line tracers, with kinematics suggestive of both bipolar and shell-like flows. These flows are thought to be driven by radio jets or shocks induced by radiation pressure originating close to the black hole, transporting dust and metal-rich material out to vast distances from the central AGN \citep{prochaska09,ivison12}. Outflows typically manifest themselves as broad wings on the line profile, and require multiple component Gaussian fits \citep[e.g.][]{nesvadba08,maiolino12}. No such features are apparent in our current data, however higher angular resolution CO mapping coupled with matched resolution radio continuum imaging could investigate whether the molecular material in our two targets is being entrained by radio jets from the AGN \citep[e.g.][]{emonts14}. We note again the tentative detection of two components in the EVN imaging of HATLAS\,J140930.4+003803 presented by \citet{virdee13}, possibly indicative of a jet.

\subsection{Luminosities, molecular gas masses, and dust masses}
\label{sec:lco}

The luminosities of the CO line in K~km~s$^{-1}$~pc$^{2}$ can be determined by the following relationship, as derived by \citet{solomon05}:	
\begin{equation}
L'_{\mathrm{CO}}~=~3.25\times10^{7}~I_{\mathrm{CO}}~\nu_{\mathrm{centre}}^{-2}~D_{\mathrm{L}}^{2}~(1+z_\mathrm{CO})^{-3}
\end{equation}
where $I_{\mathrm{CO}}$ is the velocity-integrated CO line flux in Jy~km~s$^{-1}$, $\nu_{\mathrm{centre}}$ is the observing frequency in GHz, $D_{\mathrm{L}}$ is the luminosity distance in Mpc and $z$ is the redshift of the line. CO line luminosities were derived on a per-component basis as listed in Table \ref{tab:targets}. We assume that the total CO line luminosity is the sum of the two components, also listed in the table.

Molecular ($\mathrm{H_{2}}$) gas masses are derived from $L'_{\mathrm{CO}}$ via the conversion factor $\alpha_{\mathrm{CO}}$, which has units of M$_{\odot}$~(K~km~s$^{-1}$~pc$^{2}$)$^{-1}$ \citep[for a review see][]{bolatto13}. Determining the value of $\alpha_{\mathrm{CO}}$ is an active area of research, and the choice of conversion factor is the principal source of uncertainty in high-$z$ molecular gas mass estimates. The traditional range of conversion factors ranges from 4, applicable to Giant Molecular Clouds (GMCs) in the Milky Way, to 0.8 in ULIRGs, the latter value traditionally also used for starbursting systems at high redshift \citep{downes98}. There are several theoretical models for describing $\alpha_{\mathrm{CO}}$, suggesting dependencies on gas temperature, dynamical state and metallicity \citep[e.g.][]{narayanan12}. Appropriate observational constraints allow $\alpha_{\mathrm{CO}}$ to be estimated on a per-source basis, and this has been described by several studies using dynamical or radiative transfer modelling, resolved gas surface density measurements and estimates based on dust mass. Values all tend to lie within the aforementioned 0.8--4 range: \citet{ivison11} report values in the range 0.9--2.3 for a sample of SMGs, the colour-selected galaxies studied by \citet{daddi10} have conversion factors in the range 3.6~$\pm$~0.8, and \citet{genzel12} derive values of 1.7~$\pm$~0.4 for a sample of high-$z$ star forming galaxies, corroborating the strong metallicity dependence predicted by theoretical modelling.

Adopting the standard SMG value of $\alpha_{\mathrm{CO}}$~=~0.8 M$_{\odot}$~(K~km~s$^{-1}$~pc$^{2}$)$^{-1}$ results in the molecular gas masses ($M_{\mathrm{H_{2}}}$) derived from the CO luminosities listed in Table \ref{tab:targets}, for each component and the total for each galaxy. The total H$_{2}$ masses place HATLAS\,J090426.9+015448 and HATLAS\,J140930.4+003803 amongst the most massive high-$z$ systems. The limits on $M_{\mathrm{dyn}}$ derived in Section \ref{sec:velocity} are not at odds with the $M_{\mathrm{H_{2}}}$ values, in that the latter does not exceed the former, although we emphasise again the loose constraints that we can place on $M_{\mathrm{dyn}}$  with the current data. The molecular gas in a typical high-$z$ SMG tends to be a significant fraction of the dynamical mass\footnote{Determining the ratio $M_{\mathrm{H_{2}}}$/$M_{\mathrm{dyn}}$ $>$ 1 was early evidence that the GMC conversion factor was inappropriate for high-$z$ systems.}; \citet{tacconi06,tacconi08} find fractions in the range 20--60\%. Higher resolution observations (with the VLA, ALMA or an extended ATCA configuration) would provide tighter constraints on the dynamical masses of HATLAS\,J090426.9+015448 and HATLAS\,J140930.4+003803, as well determine whether each target was a single large disk or a merger of multiple components. Such observations would then unlock independent estimates of $\alpha_{\mathrm{CO}}$ \citep[e.g.][]{ivison13}.

It is informative to compare $L'_{\mathrm{CO}}$ to the far-infrared luminosity $L_{\mathrm{FIR}}$, the former being a proxy for the star-formation potential of a galaxy and the latter being a good indicator of the star formation rate. We estimated the far infrared properties of our sample as follows. Firstly, we fit both galaxies using an isothermal modified black body model \citep[e.g.][]{hildebrand83} with fixed emissivity index \citep[$\beta$~=~1.82, following][]{smith13} and accounted for the \emph{Herschel} response curves in the 100, 160, 250, 350 and 500\,$\mu$m bands. We fit the model to the photometry on a grid of temperatures between 10 and 60~K, recording the best-fitting luminosity each time. To estimate the associated uncertainties, we created 500 Monte Carlo realisations of each galaxy by varying the photometry within the errors, and used half the difference between the 16th and 84th percentiles of the resulting luminosity distribution for each galaxy to estimate the associated uncertainties. 

HATLAS\,J090426.9+015448 and HATLAS\,J140930.4+003803 are shown on the $L'_{\mathrm{CO}}-L_{\mathrm{FIR}}$ diagram in Figure \ref{fig:lfir_lco}, showing the $L_{\mathrm{FIR}}$ values, (11.7~$\pm$1.5)~$\times$~10$^{12}$ L$_{\odot}$ and (7.6~$\pm$1.2)~$\times$~10$^{12}$ L$_{\odot}$ for HATLAS\,J090426.9+015448 and HATLAS\,J140930.4+003803 respectively, as listed in Table \ref{tab:targets}. The targets are marked by crosses on Figure \ref{fig:lfir_lco}, the extent of which show the 1$\sigma$ error bars in $L'_{\mathrm{CO}}$ and $L_{\mathrm{FIR}}$. The two sources are placed in the context of existing observations of radio galaxies (inverted triangles) as well as populations of other source types as indicated in the figure legend. These values are derived from various studies as collated by \citet{heywood13}, with additional HzRG points from \citet{emonts14}. The diagonal line shows the fit to the SMG and ultra-luminous infrared galaxy (ULIRG) population by \citet{bothwell13}. Note that some authors have claimed that this distribution features two distinct populations of galaxies, namely a `main sequence' group and a starbursting group, although others have strongly disputed this \citep{ivison11}. We refer the reader to \citet{carilli13} for further discussion.

It is clear from their position on this diagram that for their given far-infrared luminosities HATLAS\,J090426.9+015448 and HATLAS\,J140930.4+003803 have very high CO luminosities both in terms of the general distribution, but in particular for the radio galaxy population. The star formation efficiencies (SFEs) for the targets, defined as 
\begin{equation}
\mathrm{SFE}~=~\frac{L_{\mathrm{FIR}}}{L'_{\mathrm{CO}}},
\end{equation}
are 1.74~$\pm$~0.09 and 1.81~$\pm$~0.12 L$_{\odot}$ (K~km~s$^{-1}$~pc$^{2}$)$^{-1}$. This places them at the lower end of the range for SMGs, QSOs and HzRGs in the distribution shown on Figure \ref{fig:sfe_lfir}, which shows the SFE against $L_{\mathrm{FIR}}$ for the same galaxy populations plotted on Figure \ref{fig:lfir_lco}. 

The FIR modelling can also be used to provide estimates of the SFR, albeit with an unknown and un-modelled contribution to the far-infrared luminosity provided by AGN heating. Using the \citet{kennicutt98} scaling and a \citet{chabrier03} initial mass function yields 1250~$\pm$~160 M$_{\odot}$~yr$^{-1}$ and 810~$\pm$~130 M$_{\odot}$~yr$^{-1}$ for HATLAS\,J090426.9+015448 and HATLAS\,J140930.4+003803 respectively, as listed in Table \ref{tab:targets}. 

The slope of the far-infrared / radio correlation (FIRC) is captured by the dimensionless luminosity ratio 
\begin{equation}
q_{\mathrm{IR}}~=~\mathrm{log_{10}}\left[\frac{(S_{\mathrm{IR}} / 3.75~\times~10^{12})}{S_\mathrm{1.4GHz}}\right]
\end{equation}
where $S_{\mathrm{IR}}$ is the FIR flux integrated over rest-frame wavelenghts of 8--1000 $\mu$m in units of W~m$^{-2}$ and normalised to a frequency of 3.75~$\times$~10$^{12}$ Hz, and $S_{\mathrm{1.4GHz}}$ is the monochromatic, $k$-corrected radio flux density in units of W~m$^{-2}$~Hz$^{-1}$. This definition follows e.g.~\citet{ivison10a}, who report a median $q_{\mathrm{IR}}$ value of 2.44~$\pm$~0.02 for a sample of radio-selected galaxies out to $z$~$\sim$~3. Similar values are reported in numerous studies for various selection methods \citep[e.g.][]{jarvis10,bourne11,mao11,magnelli14}, often reporting moderate evolution of the FIRC with redshift.  We compute $q_{\mathrm{IR}}$ for our two targets using the properties listed in Table \ref{tab:targets}, and $k$-correcting $S_{\mathrm{1.4GHz}}$ using spectral indices derived from the 327~MHz and 1.4 GHz measurements, and the mean of the CO redshifts. This leads to $q_{\mathrm{IR}}$~=~1.92~$\pm$~0.13 and 0.81~$\pm$~0.10 for HATLAS\,J090426.9+015448 and HATLAS\,J140930.4+003803 respectively. These lie somewhat below the typical FIRC, particularly HATLAS\,J140930.4+003803, likely due to an excess of radio emission that is driven by the AGN rather than by star formation.

Since the isothermal models employed in the far-infrared modelling above do not include a contribution to the dust luminosity from very small grains which dominate at mid-infrared wavelengths, we estimate the dust luminosities using an M82 template from \citet{polletta07}, which is more luminous by around 0.36 dex than the best fit isothermal estimates. We therefore apply this correction to the isothermal dust luminosities and derive $L_{\mathrm{dust}}$~= (11.5~$\pm$~1.6)~$\times$~10$^{12}$ $L_{\odot}$ for HATLAS\,J090426.9+015448 and (7.9~$\pm$~1.1)~$\times$~10$^{12}$ $L_{\odot}$ for HATLAS\,J140930.4+003803. For the purpose of calculating dust masses, the dust emissivity is normalized at $\kappa_{850\mu m}$~=~0.77~g$^{-1}$~cm$^{2}$ \citep{dunne00}, and we use the best-fit isothermal model to each of the 500 realisations of the galaxy photometry to estimate the dust mass, with uncertainties again derived according to half the difference between the 16th and 84th percentiles of the resulting distribution. We estimate the total dust mass in each system to be (2.3~$\pm$~1.1)~$\times$~10$^{9}$~M$_{\odot}$ and (1.3~$\pm$~0.7)~$\times$~10$^{9}$~M$_{\odot}$ for HATLAS\,J090426.9+015448 and HATLAS\,J140930.4+003803 respectively. 	

These dust masses are high, comparable to the highest dust masses reported in SMGs \citep{michalowski10,rowlands14}. The typical SMG value for $\alpha_{\mathrm{CO}}$ is justifiably applicable to our targets. Combining the SFR values listed in Table \ref{tab:targets} with the molecular gas masses derived above yield gas depletion timescales of 135~$\pm$~29 Myr and 117~$\pm$~35 Myr for HATLAS\,J090426.9+015448 and HATLAS\,J140930.4+003803 respectively, assuming the star formation rate is sustained. This places the targets at the upper end of the depletion timescale distribution for QSOs, HzRGs and SMGs \citep[e.g.][]{tacconi08,emonts14,jones16}.

\begin{figure}
\centering
\includegraphics[width=\columnwidth]{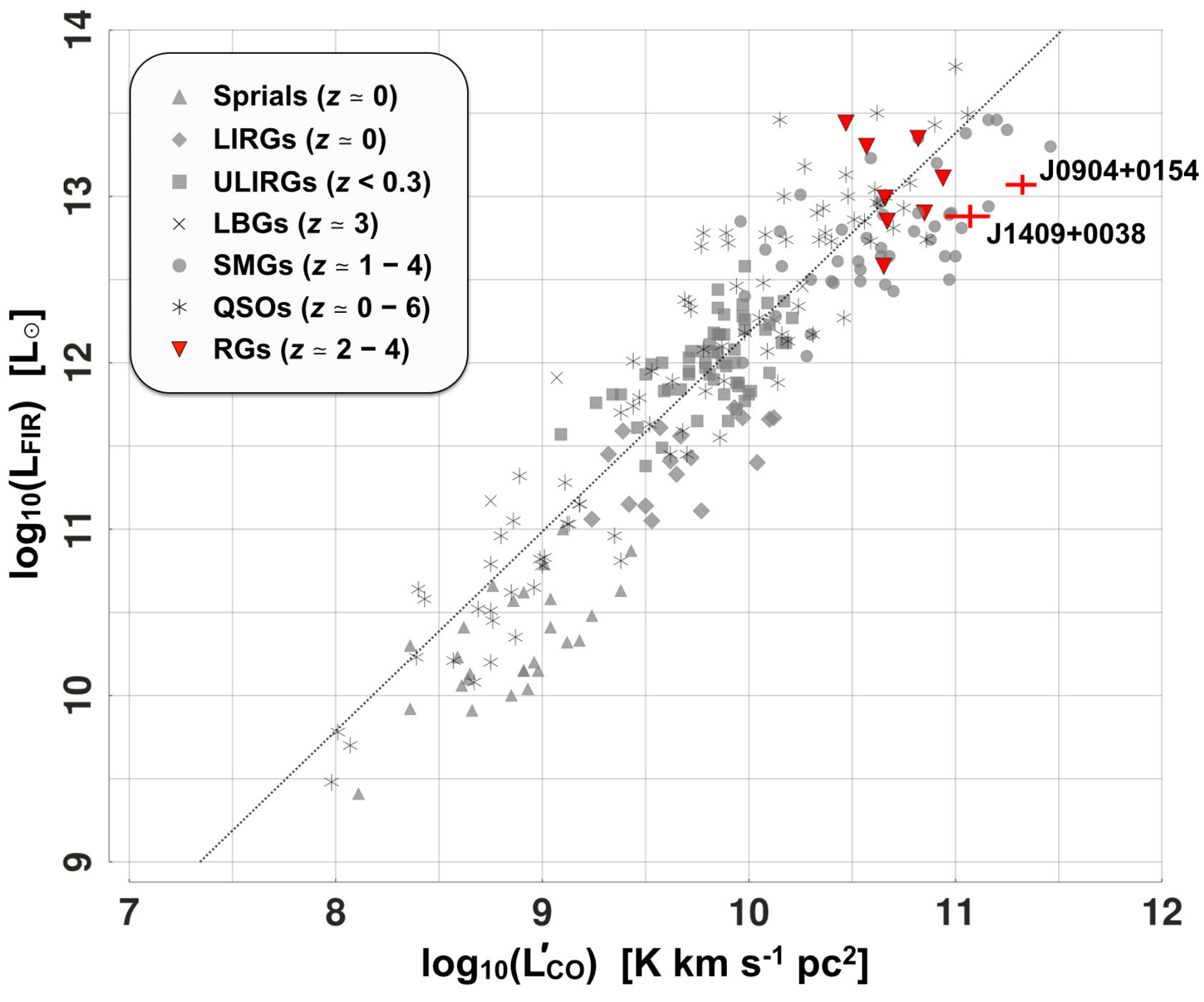}
\caption{Far infared luminosity ($L_{\mathrm{FIR}}$) against CO luminosity ($L'_{\mathrm{CO}}$) for HATLAS\,J090426.9+015448 and HATLAS\,J140930.4+003803 (labelled crosses showing the uncertainties in the two parameters), in the context of a range of sources. A sample of CO-detected radio galaxies (RGs) is shown, as well as spiral galaxies, luminous infrared galaxies (LIRGs), ultra-luminous infrared galaxies (ULIRGs), Lyman break galaxies (LBGs), sub-millimetre galaxies (SMGs) and quasars (QSOs). The diagonal line shows the fit to the SMG and ULIRG population by \citet{bothwell13}. Measurements on this plot are from various studies as collated by \citet{heywood13}, and include additional HzRG points from \citet{emonts14}.}
\label{fig:lfir_lco}
\centering
\includegraphics[width=\columnwidth]{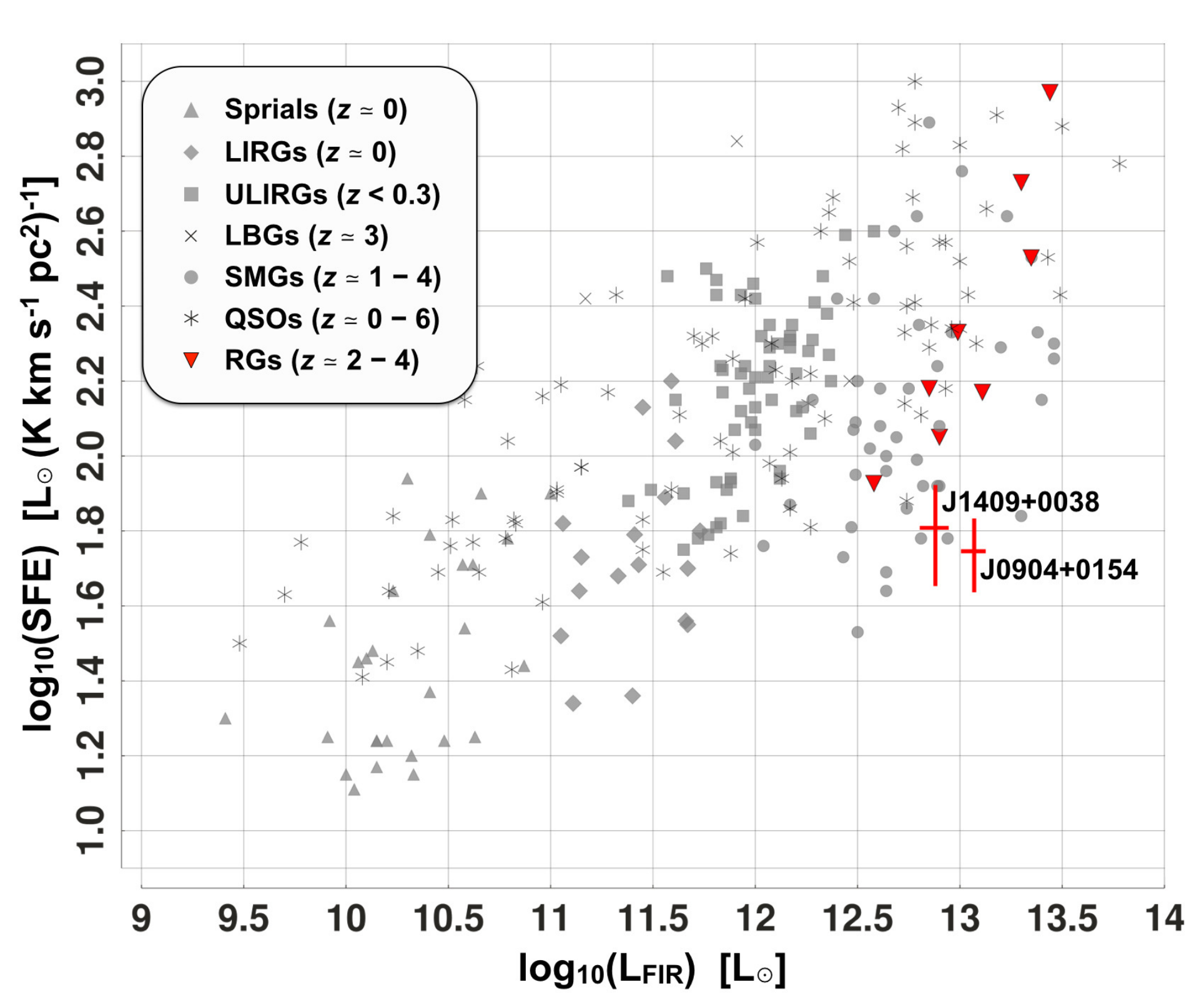}
\caption{Star formation efficiency ($L_{\mathrm{FIR}}$/$L'_{\mathrm{CO}}$) against far-infrared luminosity ($L_{\mathrm{FIR}}$) for HATLAS\,J090426.9+015448 and HATLAS\,J140930.4+003803 (labelled crosses showing the uncertainties in the two parameters). Additional points are as per Figure \ref{fig:lfir_lco}.}
\label{fig:sfe_lfir}
\end{figure}

\section{Conclusions}

We have used the Australia Telescope Compact Array to observe the CO ground-state ($J$~=~1~$\rightarrow$~0) spectra of two high-redshift radio galaxies (HzRGs). In the context of existing studies of HzRGs, our targets have relatively low 1.4 GHz radio luminosities, are close to the break in the luminosity function, and therefore represent typical systems that are undergoing high rates of accretion at high redshift. Despite this, our targets are amongst the most CO-luminous systems yet observed, with CO luminosities in excess of 10$^{11}$ K~km~s$^{-1}$~pc$^{2}$, and total inferred molecular gas masses of $\sim$10$^{11}$~M$_{\odot}$. The targets have high far-infrared luminosities, with detections in all five \emph{Herschel} bands. SED fitting to the \emph{Herschel} photometry yields far-infrared luminosities of $\sim$10$^{13}$~L$_{\odot}$, and dust masses of $\sim$10$^{9}$~M$_{\odot}$. These systems not only harbour an AGN, but share many characteristics of sub-mm galaxies, with large amounts of dust obscuring a star formation event that is producing stars at a rate of $\sim$1000~M$_{\odot}$~yr$^{-1}$. If sustained this would deplete the molecular gas reservoirs in $\sim$100 Myr. The resolved line spectra both show double-peaked profiles with total velocity ranges of $\sim$1000~km~s$^{-1}$, indicative of either an on-going merger, or a very large rotating disk of molecular material surrounding the AGN. HzRGs are thought to be the progenitors of the most massive present-day elliptical galaxies, and the merger scenario is consistent with the evolutionary picture for such objects. Existing observations with superior angular resolution tend to resolve similar systems into multiple components, and this, along with the upper limits to the dynamical masses implied by a single-disk scenario, suggests the merger interpretation is most likely to be the correct one. Molecular line observations with higher angular resolution (ideally coupled with high resolution radio continuum maps to determine the orientation of the molecular gas with respect to any radio jets) would reveal the true dynamical state of these systems. An extended ATCA configuration, the VLA or ALMA (or some combination thereof) are all viable choices of instrument for such observations. The radio continuum spectra of our targets are intriguing, peaking somewhere below 1~GHz (rest-frame) with a break in the spectrum at high radio frequencies that gives rise to a very steep spectral decline. The spectral shapes are broadly consistent with the objects being faint CSS sources, however the extreme steepness of the high-frequency radio spectrum may be caused by insufficient angular resolution of the existing radio continuum observations.

\section*{Acknowledgements}
\addcontentsline{toc}{section}{Acknowledgements}

We thank the anonymous referee and the MNRAS editorial staff for their useful comments on this paper. The Australia Telescope is funded by the Commonwealth of Australia for operation as a National Facility managed by CSIRO. The \emph{Herschel}-ATLAS is a project with \emph{Herschel}, which is an ESA space observatory with science instruments provided by European-led Principal Investigator consortia and with important participation from NASA. The H-ATLAS website is {\tt \href{http://www.h-atlas.org/}{http://www.h-atlas.org/}}. This research has made use of NASA's Astrophysics Data System. This research made use of Montage. It is funded by the National Science Foundation under Grant Number ACI-1440620, and was previously funded by the National Aeronautics and Space Administration's Earth Science Technology Office, Computation Technologies Project, under Cooperative Agreement Number NCC5-626 between NASA and the California Institute of Technology. Some figures in this paper have made use of Toyplot, an open-source plotting package for Python hosted at {\tt \href{http://toyplot.readthedocs.org/}{http://toyplot.readthedocs.org/}}. Some figures in this paper have also made use of APLpy, an open-source plotting package for Python hosted at {\tt \href{http://aplpy.github.com}{http://aplpy.github.com}}. RJI and LD acknowledge ERC Advanced Grant, COSMICISM, 321302. IH thanks Shari Breen for useful discussions.










\bsp	
\label{lastpage}
\end{document}